\documentstyle[12pt,psfig]{article}

\catcode`\@=11
\long\def\@makefntext#1{
\protect\noindent \hbox to 3.2pt {\hskip-.9pt  
$^{{\ninerm\@thefnmark}}$\hfil}#1\hfill}		

\def\@makefnmark{\hbox to 0pt{$^{\@thefnmark}$\hss}}  
	
\def\ps@myheadings{\let\@mkboth\@gobbletwo
\def\@oddhead{\hbox{}
\rightmark\hfil\ninerm\thepage}   
\def\@oddfoot{}\def\@evenhead{\ninerm\thepage\hfil
\leftmark\hbox{}}\def\@evenfoot{}
\def\sectionmark##1{}\def\subsectionmark##1{}}

\renewcommand{\thefootnote}{\fnsymbol{footnote}}

\newcounter{sectionc}\newcounter{subsectionc}\newcounter{subsubsectionc}
\renewcommand{\section}[1] {\vspace*{0.6cm}\addtocounter{sectionc}{1} 
\setcounter{subsectionc}{0}\setcounter{subsubsectionc}{0}\noindent 
	{\normalsize\bf\thesectionc. #1}\par\vspace*{0.4cm}}
\renewcommand{\subsection}[1] {\vspace*{0.6cm}\addtocounter{subsectionc}{1} 
	\setcounter{subsubsectionc}{0}\noindent 
	{\normalsize\it\thesectionc.\thesubsectionc. #1}\par\vspace*{0.4cm}}
\renewcommand{\subsubsection}[1]
{\vspace*{0.6cm}\addtocounter{subsubsectionc}{1}
	\noindent {\normalsize\rm\thesectionc.\thesubsectionc.\thesubsubsectionc. 
	#1}\par\vspace*{0.4cm}}

\newcounter{appendixc}
\newcounter{subappendixc}[appendixc]
\newcounter{subsubappendixc}[subappendixc]

\renewcommand{\appendix}[1] {\vspace*{0.6cm}
        \refstepcounter{appendixc}
        \setcounter{figure}{0}
        \setcounter{table}{0}
        \setcounter{equation}{0}
        \renewcommand{\thefigure}{\Alph{appendixc}.\arabic{figure}}
        \renewcommand{\thetable}{\Alph{appendixc}.\arabic{table}}
        \renewcommand{\theappendixc}{\Alph{appendixc}}
        \renewcommand{\theequation}{\Alph{appendixc}.\arabic{equation}}
        \noindent{\bf Appendix \theappendixc #1}\par\vspace*{0.4cm}}

\def\abstracts#1{{
	\centering{\begin{minipage}{12.2truecm}\footnotesize\baselineskip=12pt\noindent
	\centerline{\footnotesize ABSTRACT}\vspace*{0.3cm}
	\parindent=0pt #1
	\end{minipage}}\par}} 

\renewenvironment{thebibliography}[1]
	{\begin{list}{\arabic{enumi}.}
	{\usecounter{enumi}\setlength{\parsep}{0pt}
\setlength{\leftmargin 1.25cm}{\rightmargin 0pt}
	 \setlength{\itemsep}{0pt} \settowidth
	{\labelwidth}{#1.}\sloppy}}{\end{list}}

\newcounter{itemlistc}
\newcounter{romanlistc}
\newcounter{alphlistc}
\newcounter{arabiclistc}

\newcommand{\fcaption}[1]{
        \refstepcounter{figure}
        \setbox\@tempboxa = \hbox{\footnotesize Fig.~\thefigure. #1}
        \ifdim \wd\@tempboxa > 6in
           {\begin{center}
        \parbox{6in}{\footnotesize\baselineskip=12pt Fig.~\thefigure. #1}
            \end{center}}
        \else
             {\begin{center}
             {\footnotesize Fig.~\thefigure. #1}
              \end{center}}
        \fi}

\newcommand{\tcaption}[1]{
        \refstepcounter{table}
        \setbox\@tempboxa = \hbox{\footnotesize Table~\thetable. #1}
        \ifdim \wd\@tempboxa > 6in
           {\begin{center}
        \parbox{6in}{\footnotesize\baselineskip=12pt Table~\thetable. #1}
            \end{center}}
        \else
             {\begin{center}
             {\footnotesize Table~\thetable. #1}
              \end{center}}
        \fi}

\def\@citex[#1]#2{\if@filesw\immediate\write\@auxout
	{\string\citation{#2}}\fi
\def\@citea{}\@cite{\@for\@citeb:=#2\do
	{\@citea\def\@citea{,}\@ifundefined
	{b@\@citeb}{{\bf ?}\@warning
	{Citation `\@citeb' on page \thepage \space undefined}}
	{\csname b@\@citeb\endcsname}}}{#1}}

\newif\if@cghi
\def\cite{\@cghitrue\@ifnextchar [{\@tempswatrue
	\@citex}{\@tempswafalse\@citex[]}}
\def\citelow{\@cghifalse\@ifnextchar [{\@tempswatrue
	\@citex}{\@tempswafalse\@citex[]}}
\def\@cite#1#2{{$\null^{#1}$\if@tempswa\typeout
	{IJCGA warning: optional citation argument 
	ignored: `#2'} \fi}}

 1
 1
 1

\font\ninerm=cmr9

\begin{document}

\noindent PM--97--01  \hspace*{11cm} February 1997 

\bigskip

\centerline{\normalsize\bf SUSY HIGGS BOSON DECAYS\footnote{Talk 
given at the Ringberg Workshop {\it the Higgs Puzzle}, Ringberg Castle, 
Tegernsee, Germany, December 8--13 1996; to appear in the proceedings.}}
\baselineskip=16pt
\centerline{\footnotesize ABDELHAK DJOUADI}
\baselineskip=13pt
\centerline{\footnotesize\it 
Laboratoire de Physique Math\'ematique et Th\'eorique, UPRES--A 5032, }
\baselineskip=12pt
\centerline{\footnotesize\it 
Universit\'e de Montpellier II, F--34095 Montpellier Cedex 5, France.}
\centerline{\footnotesize E-mail: djouadi@lpm.univ-montp2.fr}
\vspace*{0.3cm}
\vspace*{0.9cm}
\abstracts{
I discuss the decay modes of the neutral and charged Higgs bosons
in the minimal supersymmetric extension of the Standard Model. Special emphasis
will be put on the the QCD corrections to the hadronic decay modes, the 
below threshold --three body-- decays and the decays into supersymmetric 
particles, charginos, neutralinos and sfermions. A Fortran code calculating the 
various Higgs decay branching ratios is then briefly presented.} 
 
\normalsize\baselineskip=15pt
\setcounter{footnote}{0}
\renewcommand{\thefootnote}{\alph{footnote}}


\newcommand{\lsim}{\raisebox{-0.13cm}{~\shortstack{$<$ \\[-0.07cm] $\sim$}}~}
\newcommand{\gsim}{\raisebox{-0.13cm}{~\shortstack{$>$ \\[-0.07cm] $\sim$}}~}
\newcommand{\ra}{\rightarrow}
\newcommand{\s}{\\ \vspace*{-3mm} }
\newcommand{\nn}{\noindent}
\newcommand{\non}{\nonumber}
\newcommand{\beq}{\begin{eqnarray}}
\newcommand{\eeq}{\end{eqnarray}}
\newcommand{\tb}{\mbox{tg}\beta}
\newcommand{\GeV}{\rm GeV}

\section{Introduction}

In the Minimal Supersymmetric extension of the Standard Model$^1$ (MSSM), the
Higgs sector$^2$ is extended to comprise three neutral $h/H$ (CP=+), $A$ 
(CP=--) and a pair of charged scalar particles $H^\pm$. The Higgs sector is 
highly constrained since there are only two free parameters at tree--level: 
a Higgs mass parameter [generally $M_A$] and the ratio of the vacuum 
expectation values of the two doublet fields responsible for the symmetry 
breaking, $\tb$ [which in Grand Unified Supersymmetric models with $b$--$\tau$ 
Yukawa coupling unification$^3$ is forced to be either small, $\tb \sim 1.5$, 
or large, $\tb \sim 50$]. After the inclusion of the large radiative 
corrections$^4$, while the lightest Higgs boson $h$ is predicted to be lighter 
than $M_h \lsim 130$ GeV, the $H,A$ and $H^\pm$ states are expected to 
have masses of the order of a few hundred GeV.

The decay pattern of the MSSM Higgs bosons is determined to a large extent 
by their couplings to fermions and gauge bosons, which in general depend 
strongly on $\tb$ and the mixing angle $\alpha$ in the CP--even sector. 
The pseudoscalar and charged Higgs boson couplings to down (up) type fermions 
are (inversely) proportional to $\tb$; the pseudoscalar $A$ has no tree level 
couplings to gauge bosons.  For the CP--even Higgs 
bosons, the couplings to down (up) type fermions are enhanced (suppressed) 
compared to the SM Higgs couplings [$\tb>1$]; the couplings to gauge bosons 
are suppressed by $\sin/\cos(\beta-\alpha)$ factors [see Table 1.]

For large values of $\tb$ the pattern is simple, a result of the strong
enhancement of the Higgs couplings  to down--type fermions. The neutral
Higgs bosons will decay into $b\bar{b}$ ($\sim 90\%$) and $\tau^+ \tau^-$ 
($\sim 10\%)$ pairs, and $H^\pm$ into $\tau \nu_\tau$ pairs below and $tb$ 
pairs above the top--bottom threshold. 
For the CP--even Higgs bosons $h$ and $H$, only when $M_h$ 
approaches its maximal value is this simple rule modified: in this decoupling 
limit, the $h$ boson is SM--like and decays into charm and gluons with a
rate similar to the one for $\tau^+ \tau^- $ [$\sim 5\%$] and in the high 
mass range, $M_h \sim 130$ GeV, into $W$ pairs with one of the $W$  bosons
being virtual; the $H$ boson will mainly decay into $hh$ and $AA$ final 
states.

\bigskip

\begin{center}
\begin{tabular}{|c|c|c|c|c|} \hline
$\ \ \ \Phi \ \ \ $ &$ g_{\Phi \bar{u}u} $      & $ g_{\Phi \bar{d} d} $ &
$g_{ \Phi VV} $ \\ \hline
$h$  & \ $\; \cos\alpha/\sin\beta \rightarrow 1     \; $ \ & \ $ \; -\sin\alpha/
\cos\beta \rightarrow 1 \; $ \ & \ $ \; \sin(\beta-\alpha) \rightarrow 1 \; 
$ \ \\
 $H$  & \   $\; \sin\alpha/\sin\beta \rightarrow 1/\tb \; $ \ & \ $ \; 
\cos\alpha/ \cos\beta \rightarrow \tb \; $ \ & \ $ \; \cos(\beta-\alpha) 
\rightarrow 0 \; $ 
\ \\ $A$  & \ $\; 1/ \tb \; $\ & \ $ \; \tb \; $ \   & \ $ \; 0 \; $ \ \\ \hline
\end{tabular}
\end{center}
\vspace{0.3cm}
\nn {\small Table 1: Higgs couplings to fermions and gauge bosons normalized
to the SM Higgs couplings, and their limit for $M_A \gg M_Z$.}

\bigskip 

For small values of $\tb\sim 1$ the decay pattern of the heavy neutral
Higgs bosons is much more complicated. The $b$ decays are in general
not dominant any more; instead, cascade decays to pairs of light Higgs
bosons and mixed pairs of Higgs and gauge bosons are important and
decays to $WW/ZZ$ pairs will play a role. For very large masses, they decay 
almost exclusively to top quark pairs. The decay pattern of the charged 
Higgs bosons for small $\tb$ is similar to that at large $\tb$ except in 
the intermediate mass range where cascade decays to $Wh$ are dominant.

When the decays into supersymmetric particles are kinematically allowed, as 
it should be the case at least for the heavy CP--even, CP--odd and charged 
Higgs bosons, the pattern becomes even more complicated since the
decay channels into charginos, neutralinos and squarks play will play 
a non--negligible role.   

In the following, I will discuss three topics related to the decay modes 
of the Higgs particles in the MSSM: (a) the QCD corrections to the hadronic 
decay modes$^5$, (b) the below threshold three--body decays$^6$ and (c) the 
decays into SUSY particles$^{7}$ of the heavy $H,A$ and $H^\pm$ bosons,
including the QCD corrections to the squark decay modes$^8$.  I will then 
briefly introduce a Fortran code$^9$ which calculates the various decay 
branching ratios. For more details and for a complete list of references, 
see the original papers Refs.~[5--9].

\section{Hadronic Decay Modes: QCD Corrections$^5$} 

The particle width for decays to massless $b,c$ quarks directly coupled to
the Higgs particle is given, up to ${\cal O}(\alpha_{s}^{2})$ QCD 
corrections [the effect of the electroweak radiative corrections in the 
branching ratios is negligible], by the well-known expression
\begin{small}
\begin{eqnarray}
\Gamma [\Phi  \ra Q{\overline{Q}}] = \frac{3 G_F M_\Phi } {4\sqrt{2}\pi}
g^2_{\Phi QQ} \overline{m}_Q^2(M_\Phi) \left[ 1 + 5.67 \frac{\alpha_s} 
{\pi} + (35.94 - 1.36 N_F) \frac{\alpha_s^2}{\pi^2} \right] 
\end{eqnarray}
\end{small}
in the ${\overline{\rm MS}}$ renormalization scheme; the running quark
mass and the QCD coupling are defined at the scale of the Higgs mass,
absorbing this way any large logarithms. The quark masses can be neglected
in general except for top quark decays where this approximation holds only 
sufficiently far above threshold. Since the relation between the charm 
pole mass $M_{c}$ and the ${\overline{\rm MS}}$ mass evaluated 
at the pole mass ${\overline{m}}_{c}(M_{c})$ is badly convergent, one can 
adopt the running quark masses ${\overline{m}}_{Q}(M_{Q})$ [which have been 
extracted directly from QCD sum rules evaluated in a consistent ${\cal O}
(\alpha_{s})$ expansion] as starting points. The evolution from $M_{Q}$ 
to a scale $\mu \sim M_\Phi$ is given by:
\begin{eqnarray}
{\overline{m}}_{Q} (\mu )={\overline{m}}_{Q}\,(M_{Q})
\, c[\alpha_{s}(\mu)/\pi ] / c [\alpha_{s}(M_{Q})/\pi]   \nonumber 
\end{eqnarray}
\begin{eqnarray} 
c(x) &=& (25/6x)^{12/25} \, [1+1.014x+1.39x^{2}] \ \ {\rm for} \ 
M_{c}<\mu <M_{b} \nonumber \\
c(x) &=&(23/6 x)^{12/23} \, [1+1.175x+1.50 x^{2}] \ \ {\rm for} \ M_{b}<\mu  
\end{eqnarray} 

Typical values of the running $b,c$ masses at the scale $\mu = 100$ GeV, 
characteristic for $M_\Phi$, are displayed in Table 2, with
the evolution calculated  for $\alpha_{s}(M_{Z})=0.118\,\pm \, 0.006$;
$M_{Q}^{\rm pt2}$ are the quark pole masses. 

\bigskip

\begin{center}
\begin{tabular}{|c|c|cc|c|} \hline
&$ \alpha_{s}(M_{Z}) $
& $ {\overline{m}}_{Q}(M_{Q})$\ & \ $M_{Q}\,=\,M_{Q}^{\rm pt2} $
& \ ${\overline{m}}_{Q}\,(\mu\,=\ 100~\GeV) $ \\ \hline
$b$ & $0.112$ & $(4.26 \pm 0.02)~\GeV$ & $(4.62 \pm 0.02)~\GeV$
& $(3.04 \pm 0.02)~\GeV$ \\
    & $0.118$ & $(4.23 \pm 0.02)~\GeV$ & $(4.62 \pm 0.02)~\GeV$
& $(2.92 \pm 0.02)~\GeV$ \\
    & $0.124$ & $(4.19 \pm 0.02)~\GeV$ & $(4.62 \pm 0.02)~\GeV$
& $(2.80 \pm 0.02)~\GeV$ \\ \hline 
$c$ & $0.112$ & $(1.25 \pm 0.03)~\GeV$ & $(1.42 \pm 0.03)~\GeV$
& $(0.69 \pm 0.02)~\GeV$ \\
    & $0.118$ & $(1.23 \pm 0.03)~\GeV$ & $(1.42 \pm 0.03)~\GeV$
& $(0.62 \pm 0.02)~\GeV$ \\
    & $0.124$ & $(1.19 \pm 0.03)~\GeV$ & $(1.42 \pm 0.03)~\GeV$
& $(0.53 \pm 0.02)~\GeV$ \\ \hline
\end{tabular}
\end{center}

\bigskip 

\nn {\small Table 2: The running $b$ and $c$ quark masses in the
$\overline{\rm MS}$ scheme at a scale $\mu=100$ GeV.}

\bigskip 

The decay of the Higgs bosons to gluons is, to a good approximation mediated by 
heavy top quark loops; the partial decay width, including  QCD radiative
corrections which are built up by the exchange of virtual gluons and the
splitting of a gluon into two gluons or into $N_F$ massless quark--antiquark 
pairs, is given by [$\mu \sim M_{\Phi}$]
\begin{small}
\begin{eqnarray}
\Gamma^{N_F} [ \phi \ra gg+..] &=& \frac{G_{F} g^2_{\phi tt} 
\alpha_{s}^2 M_{\phi}^{3}} {36  \sqrt{2} \pi^{3}}  
\left[ 1+ \frac{\alpha_s}{\pi} \left( \frac{95}{4} 
-\frac{7}{6}N_{F} +\frac{33-2N_{F}} {6}\log \frac{\mu ^{2}}{M_{\phi}^{2}}
 \right) \right] \non \\
\Gamma^{N_F} [ A \ra gg+..] &=& \frac{G_{F} g^2_{A tt} 
\alpha_{s}^2 M_{A}^{3}} {16  \sqrt{2} \pi^{3}}  
\left[ 1+ \frac{\alpha_s}{\pi} \left( \frac{97}{4} 
-\frac{7}{6}N_{F} +\frac{33-2N_{F}} {6}\log \frac{\mu ^{2}}{M_{A}^{2}}
 \right) \right]
\end{eqnarray}
\end{small}
with $\phi = h,H$ and $\alpha_s\equiv \alpha_s^{N_F}(\mu^2)$. The radiative 
corrections are very large, nearly doubling the partial width. The final 
states $\Phi \ra b{\overline{b}}g$ and $c{\overline{c}}g$ are also
generated through processes in which the $b,c$ quarks are coupled to the Higgs
boson directly. Gluon splitting $g\ra b{\overline{b}}$ in $\Phi \ra 
gg$ increases the inclusive decay probabilities $\Gamma(\Phi \ra 
b\bar{b}+ \dots)$ {\it etc.} Since $b$ quarks, and eventually $c$ quarks, 
can in principle be tagged experimentally, it is physically meaningful to 
consider the particle width of Higgs decays to gluon and light $u,d,s$ quark 
final jets separately. The contribution of $b,c$ quark final states to the 
coefficient in front of $\alpha_s$ in eq.~(3) is:
$$ 
-\frac{7}{3} + \frac{1}{3} [\log \frac{M_{\Phi}^{2}} {M_{b}^{2}}+\log 
\frac{M_{\Phi}^{2}} {M_{c}^{2}} ]
$$
Instead of naively subtracting this 
contribution, it may be noticed that the mass logarithms can be absorbed 
by changing the number of active flavors from $N_{F}=5$ to $N_{F}=3$ in 
the QCD coupling $\alpha_{s}^{(N_F)}$. The subtracted parts may be added to 
the partial decay widths into $c$ and $b$ quarks. 

The numerical analysis of the branching ratios for the lightest CP--even 
Higgs decays in the decoupling limit where $h$ is SM--like, with the quark 
masses and QCD couplings 
given above and a top mass $M_{t}=(176 \pm 11)$ \GeV, is shown in Fig.~1. 
To estimate systematic uncertainties, the variation of the $c$ mass has
been stretched over $2\sigma$ and the uncertainty of the $b$ mass to
0.05 \GeV. However, the dominant error in the predictions is due to the
uncertainty in $\alpha_{s}$ and the errors in the prediction for 
the charm and gluon branching ratios are very large. Nevertheless, the expected 
hierarchy of the Higgs decay modes is clearly visible despite these 
uncertainties. Similar results hold for the heavy CP--even and CP--odd 
Higgs decays. 

\begin{figure}[htbp]
\vspace*{-.7cm}
\psfig{figure=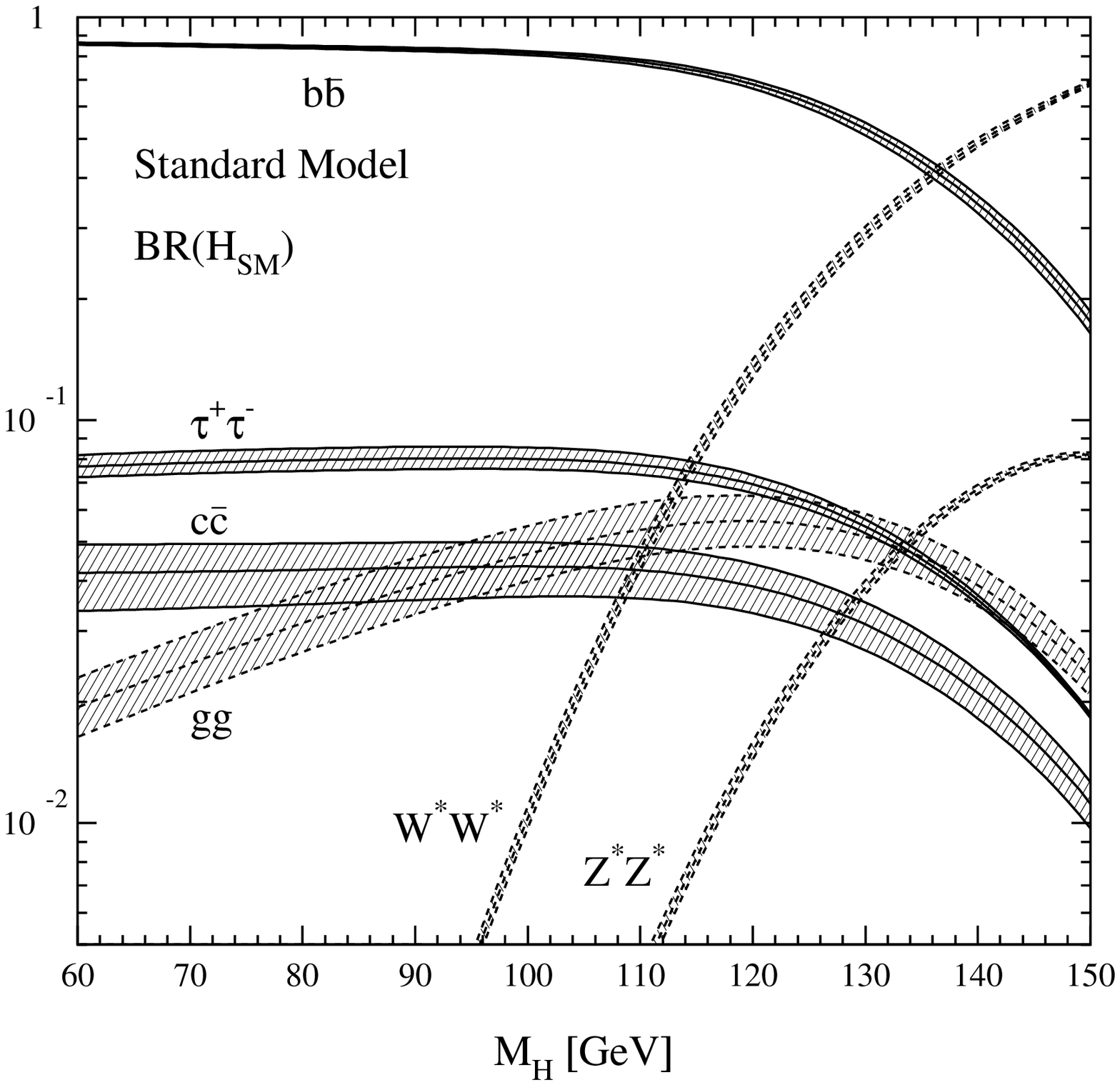,height=11.5cm,angle=-90}
\vspace*{-1cm}
\noindent {\small Fig.~1: Branching ratios of the $h$ boson in the decoupling 
limit, including the uncertainties from the quark masses and the QCD coupling 
$\alpha_s$ [shaded bands].}
\end{figure}

\pagebreak

\section{Three Body decay modes$^6$}

Besides these two--body decays, below--threshold modes can play an important 
role. It is well--known that SM Higgs decays
into real and virtual $Z$  pairs are quite substantial: the suppression
by the off--shell propagator and the additional $Zff$ coupling is at
least partly compensated by the large Higgs coupling to the $Z$ bosons.
For the same reason, three--body decays of MSSM Higgs particles mediated
by gauge bosons, heavy Higgs bosons and top quarks, are of physical
interest. Important three-body decays for the $H,A$ and $H^\pm$ bosons are
$[V=W,Z]$: 
\begin{eqnarray}
H &\ra& VV^* \ra V f \bar{f}^{(')} \ , \
        AZ^* \ra  A f \bar{f} \ , \
        H^\pm W^{\mp *} \ra  H^\pm f \bar{f}' \ , \
        \bar{t}t^* \ra \bar{t} b W^+ \\
A & \ra & hZ^* \ra h f \bar{f} \ , \
        \bar{t}t^* \ra \bar{t} bW^+  \\
H^\pm & \ra & hW^* \ra h f \bar{f}' \ , \
        AW^* \ra A f \bar{f}' \ , \
        \bar{b}t^* \ra \bar{b}bW
\end{eqnarray}
For the lightest Higgs boson $h$, the only releveant below threshold decay 
mode is $h \ra W^* W^*$ for $M_h \sim 130$ GeV. In this case, both the $W$'s 
have to be taken off--shell. The branching ratios for $h,H,A$ and $H^\pm$ 
decays are shown in Fig.~2 for $\tb=1.5$, in the case where the mixing in the 
stop sector is neglected.

For the heavy Higgs boson $H$, the decay $H\ra hh$ is the dominant
channel, superseded by $t\bar{t}$ decays above the threshold [for the
latter, the inclusion of the three--body modes provides a smooth
transition from below to above threshold]. This rule is only broken for
Higgs masses of about 140 GeV where an accidentally small value of the
$\lambda_{Hhh}$ coupling allows the $b \bar{b}$ and $WW^*$ decay modes
to become dominant. Important channels in general, below the $t\bar{t}$
threshold, are decays to pairs of gauge bosons and $b\bar{b}$ decays.
In a restricted range of $M_H$, below--threshold $AZ^*$ and
$H^{\pm}W^{\mp *}$ also play a non--negligible role.
In the case of the pseudoscalar $A$, the dominant modes are the $A\ra
b\bar{b}$ and $A \ra t\bar{t}$ decays below the $hZ$ and $t\bar{t}$
thresholds respectively; in the intermediate mass region, $M_A=200$ to
$300$ GeV, the decay $A \ra hZ^*$ [which reaches $\sim 1\%$ already at
$M_A =130$ GeV] dominates. The gluonic decays are significant around the
$t\bar{t}$ threshold.
For the charged Higgs boson, the inclusion of the three--body decay
modes will reduce the branching ratio for the $\tau\nu$ channel quite
significantly. Indeed, this decay does not overwhelm all the other modes
since the three--body decay channels $H^+ \ra hW^*$ as well as
$H^+\ra AW^*$ in the low mass range and $H^+ \ra bt^*$ in the
intermediate mass range have appreciable branching ratios.

The total widths of the Higgs bosons are in general considerably
smaller than for the SM Higgs due to the absence or the
suppression of the decays to $W/Z$ bosons which grow as $M_H^3$. 
The dominant decays are built-up by top quarks so that
the widths rise only linearly with $M_\Phi$. However, for large $\tb$ values, 
the decay widths scale in general like ${\rm tg}^2\beta$ and can 
become experimentally significant, for $\tb \gsim {\cal O}(30)$ and
for large $M_\Phi$.

\begin{figure}
\vspace*{-1.cm}
\centerline{\psfig{figure=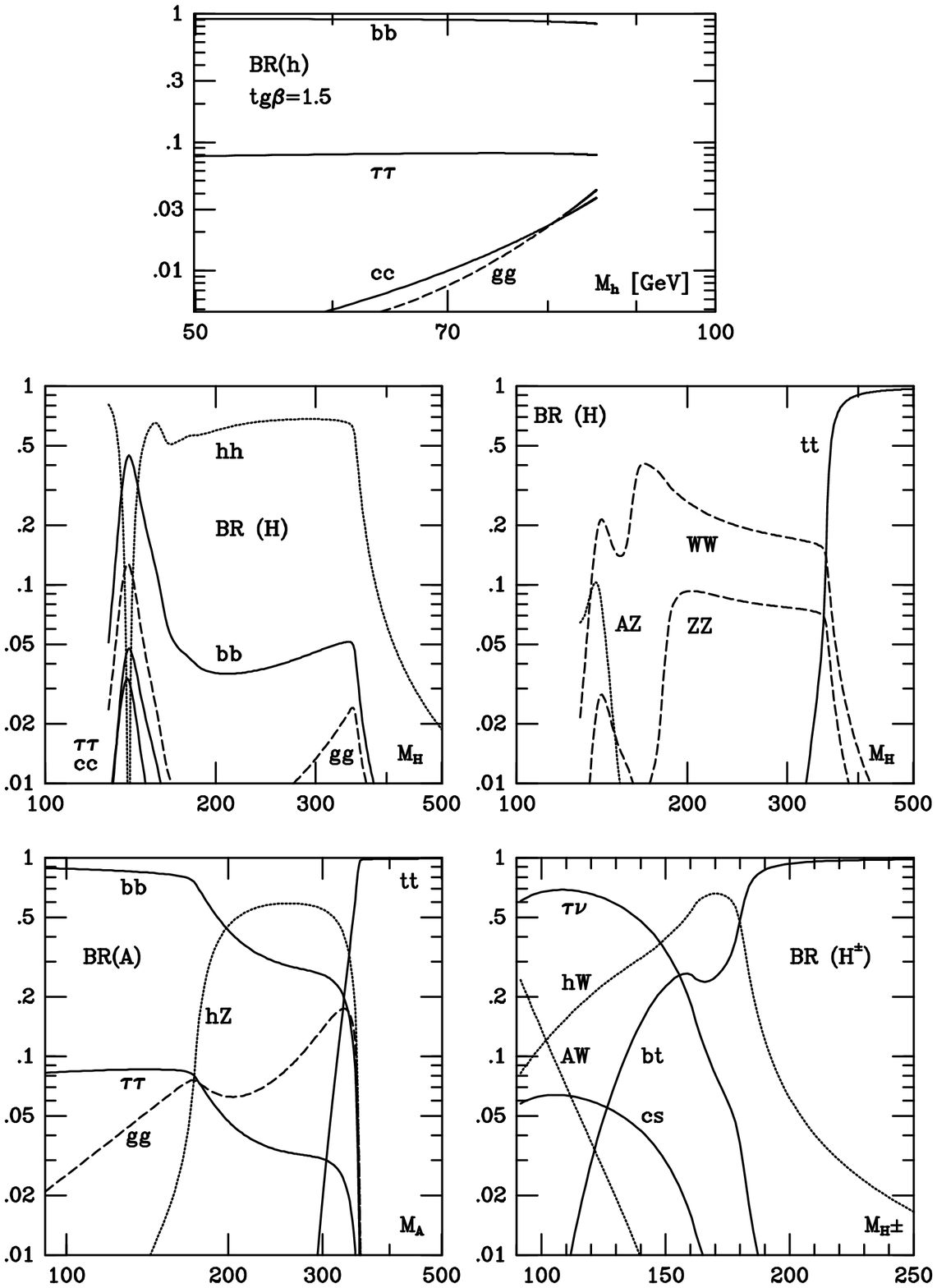,height=21cm}}
\vspace*{-2.cm}
\noindent {\small Fig.~2: Branching ratios for the CP--even, the CP--odd 
and the charged MSSM Higgs bosons, including the three--body decays, for 
$\tb=1.5$ and no stop mixing.}
\end{figure}

\section{SUSY Decay modes$^7$}

In the previous discussion, we have assumed that decay channels into
neutralinos, charginos and sfermions are shut. However, these channels
could play a significant role, since some of these particles can 
have masses in the ${\cal O}(100$ GeV) range or less.
To discuss these decays, we will restrict ourselves to the MSSM 
constrained by minimal Supergravity, in which the SUSY sector is 
described in terms of five
universal parameters at the GUT scale: the common scalar mass $m_0$, the
common gaugino mass $M_{1/2}$, the trilinear coupling $A$, the bilinear
coupling $B$ and the higgsino mass $\mu$. These parameters evolve
according to the RGEs, forming the supersymmetric particle spectrum at
low energy.

The requirement of radiative electroweak symmetry breaking further
constrains the SUSY spectrum, since the minimization of the one--loop
Higgs potential specifies the parameter $\mu$ [to within a sign] and
also $B$. The unification of the $b$ and $\tau$ Yukawa couplings gives
another constraint: in the $\lambda_t$ fixed--point region, the value of
$\tb$ is fixed by the top quark mass through: $m_t \simeq (200~{\rm
GeV}) \sin\beta$, leading to $\tb \simeq 1.75$. There also exists a
high--$\tb$ [$\lambda_b$ and $\lambda_\tau$ fixed--point] region for
which $\tb \sim$ 50--60.  If one also notes that moderate values of
the trilinear coupling $A$ have little effect on the resulting spectrum, 
then the whole SUSY spectrum will be a function of  $\tb$ which we
take to be $\tb=1.75$ and 50, the sign of $\mu$, $m_0$ which in practice we 
replace with $M_A$ taking the two illustrative values $M_A =300$ and 600 
GeV, and the common gaugino mass $M_{1/2}$ that we will freely vary.

The decay widths of the heavy CP-even, the CP--odd and the charged Higgs
bosons, $H,A$ and $H^\pm$, into pairs of neutralinos and
charginos [dashed lines], squarks [long--dashed  lines] and sleptons
[dot--dashed lines], as well as the total [solid lines] and non--SUSY
[dotted--lines] decay widths, are shown in Fig.~3 for $\tb=1.75$, $\mu>0$
and two values of $M_A=300$ [left curves] and $600$ GeV [right
curves].

For $M_A=300$ GeV, i.e. below the $t\bar{t}$ threshold, the widths of
the $H$ decays into inos and sfermions are much larger than
the non--SUSY decays. In particular, squark [in fact $\tilde{t}$ and 
$\tilde{b}$ only] decays are almost two--orders of magnitude larger when
kinematically allowed. The situation changes dramatically for larger $M_A$
when the $t\bar{t}$ channel opens up: only the decays into $\tilde{t}$ 
pairs when allowed are competitive with the dominant $H \ra t\bar{t}$
channel. Nevertheless, the decays into inos are still substantial having
BRs at the level of 20\%; the decays into sleptons never exceed a few percent.

In the case of the pseudoscalar $A$, because of CP--invariance and the
fact that sfermion mixing is small except in the stop sector, only the
decays into inos and $A \ra \tilde{t}_1 \tilde{t}_2$
decays are allowed. For these channels, the situation is quite similar
to the case of $H$: below the $t\bar{t}$ threshold the decay width into
ino pairs is much larger than the non--SUSY decay widths [here
$\tilde{t}_2$ is too heavy for the $A \ra \tilde{t}_1 \tilde{t}_2$ decay
to be allowed], but above $2m_t$ only the $A \ra \tilde{t}_1
\tilde{t}_2$ channel competes with the $t\bar{t}$ decays.

For the charged Higgs boson $H^\pm$, only the decay  $H^+ \ra
\tilde{t}_1 \tilde{b}_1$ [when kinematically allowed] competes with the
dominant  $H^+\ra t\bar{b}$ mode, yet the $\tilde{\chi}^+ \tilde{
\chi}^0$ decays have a branching ratio of a few ten percent; the
decays into sleptons are at most of ${\cal O}(\%)$. 

In the case where $\mu<0$, the situation is quite similar as above. For
large $\tb$ values, $\tb\sim 50$, all gauginos and sfermions are very
heavy and therefore kinematically inaccessible, except for the lightest
neutralino and the $\tau$ slepton. Moreover, the $b\bar{b}/\tau \tau$
and $t\bar{b}$/$\tau \nu$ [for the neutral and charged Higgs bosons
respectively] are enhanced so strongly, that they leave no chance for
the SUSY decay modes to be significant. Therefore, for large $\tb$, the
simple pattern of $bb/\tau\tau$ and $tb$ decays for heavy neutral and
charged Higgs bosons still holds true even when the SUSY decays are
allowed.

\smallskip

\begin{figure}[htbp]
\vspace*{.3cm}
\centerline{\psfig{figure=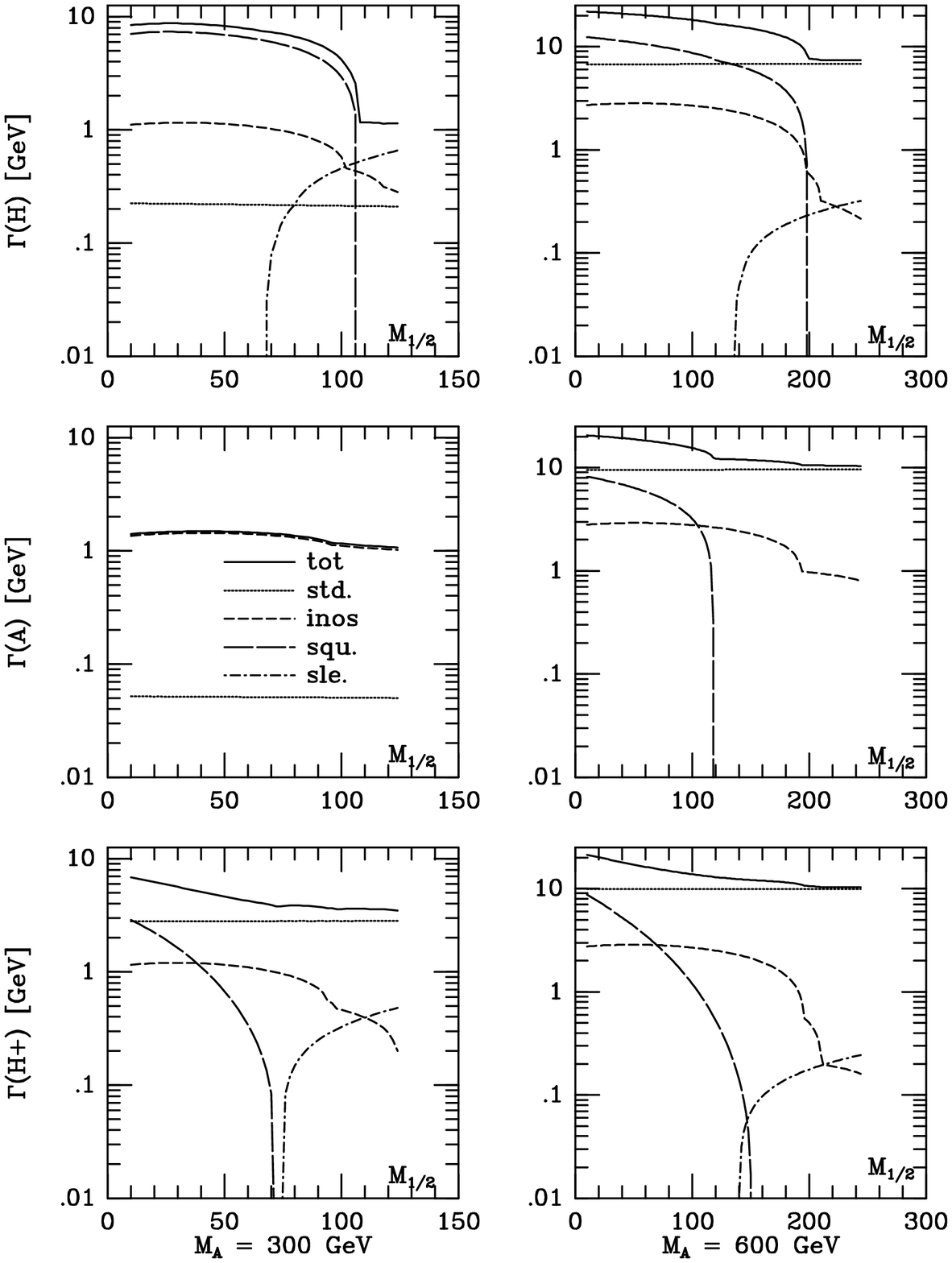,height=15cm}}
\vspace*{-1.5cm}
\noindent {\small Fig.~3: Decay widths for the SUSY decay modes of the
heavy CP--even, CP--odd and charged Higgs bosons, for $\tb=1.75$.
The total and the non--SUSY widths are also shown.}
\end{figure}

\newpage 

Since the decays into stop and sbottom squarks can be dominant when
kinematically allowed, QCD corrections must be incorporated in order to
have full control on the decay widths and to make a reliable comparison
with the standard (non--SUSY) decay channels. The QCD corrections to 
the decays of the heavy CP--even, CP--odd and charged MSSM Higgs bosons
into stop and sbottom quarks have been recently calculated$^8$. These
corrections are found to be rather large, enhancing or supressing the
widths by amounts up to 50\% and in some case even more. The QCD corrections
depend strongly on the gluino mass; however for very heavy gluinos, 
they are only logarithmically dependent on $m_{\tilde{g}}$. 
Contrary to the case of Higgs boson decays into light quark pairs, these
large corrections cannot be absorbed into running squark masses since the 
latter are expected to be of the same order as as the Higgs masses.

\section{The program HDECAY$^9$}

Finally, let me shortly describe the fortran code HDECAY, which
calculates the various decay widths and the branching ratios of 
Higgs bosons in the SM and the MSSM and which includes: 

(a) All decay channels that are kinematically allowed and which have
branching ratios larger than $10^{-4}$, {\it y compris} the loop
mediated, the three body decay modes and in the MSSM the cascade and the
supersymmetric decay channels.

(b) All relevant two-loop QCD corrections to the decays into quark pairs and 
to the quark loop mediated decays into gluons are incorporated in the most 
complete form; the small leading electroweak corrections are also included. 

(c) Double off--shell decays of the CP--even Higgs bosons into massive
gauge bosons which then decay into four massless fermions, and all 
all important below--threshold three--body decays discussed previously. 

(d) In the MSSM, the complete radiative corrections in the effective
potential approach with full mixing in the stop/sbottom sectors;
it uses the renormalisation group improved values of the Higgs masses
and couplings and the relevant leading next--to--leading--order corrections 
are also implemented. 

(e) In the MSSM, all the decays into SUSY particles (neutralinos, charginos,
sleptons and squarks including mixing in the stop, sbottom and stau 
sectors) when they are kinematically allowed. The SUSY particles are also 
included in the loop mediated $\gamma \gamma$ and $gg$ decay channels.

\smallskip

The basic input parameters, fermion and gauge boson masses and total
widths, coupling constants and in the MSSM, soft--SUSY breaking
parameters can be chosen from an input file. In this file several flags
allow to switch on/off or change some options [{\it e.g.} chose a
particular Higgs boson, include/exclude the multi--body or SUSY decays,
or include/exclude specific higher--order QCD corrections]. The results for the
many decay branching ratios and the total decay widths are written to
several output files with headers indicating the processes and giving
the input parameters. 

\newpage

\section{Acknowledgments:} 

I thank the organisers of this workshop, in particular Bernd Kniehl, for 
the nice and stimulating atmosphere of the meeting. The work discussed
here was done in enjoyable collaborations with Abdeslam Arhrib, Wolfgang
Hollik, Jan Kalinowski, Patrick Janot, Christoph J\"unger, Paul Ohmann, 
Michael Spira and Peter Zerwas.

\end{document}